\documentclass[12pt]{article}
\pdfoutput=1

\usepackage{amsmath,amssymb,amscd}
\usepackage{listings}
\usepackage{caption}
\usepackage{slashed}
\usepackage{color}

\usepackage[pdftex]{graphicx}
\usepackage{epstopdf}
\usepackage{epsfig}
\usepackage{listings}
\usepackage{caption}
\usepackage{cite}

\usepackage{multirow}

\usepackage{graphicx}
\usepackage{xspace}
\usepackage{floatrow}
\usepackage{amsmath}
\usepackage{amsfonts}
\usepackage{amssymb}
\usepackage{calrsfs}
\usepackage{wrapfig}

\usepackage{titlesec}

\titleformat*{\section}{\large\bfseries}
\let\OLDthebibliography\thebibliography
\renewcommand\thebibliography[1]{
  \OLDthebibliography{#1}
  \setlength{\parskip}{0pt}
  \setlength{\itemsep}{5pt plus 0.3ex}
}

\newcommand{\epem}{\rm e^+e^-}

\newcommand{\sqrts}{\sqrt{\rm s}}

\newcommand{\ttbar}{\rm t\overline{t}}
\newcommand{\ssbar}{\mathrm{s\bar{s}}}
\newcommand{\ddbar}{\mathrm{d\bar{d}}}
\newcommand{\uubar}{\mathrm{u\bar{u}}}

\newcommand{\Lint}{\mathcal{L}_{\mbox{\rm \tiny{int}}}}

\newcommand*{\eg}{e.g.\@\xspace}
\newcommand*{\ie}{i.e.\@\xspace}
\newcommand*{\cm}{c.m.\@\xspace}
\newcommand*{\vs}{vs.\@\xspace}

\def\cO#1{{{\cal{O}}}\left(#1\right)}

\newfloatcommand{capbtabbox}{table}[][\FBwidth]

\begin{document}

\begin{titlepage}

\pagestyle{empty}

\baselineskip=21pt
{\small
\rightline{38th Intl Conf. on High Energy Physics, 3--10 Aug. 2016, Chicago (USA)}
\rightline{Proceds. ICHEP'16, \bf{PoS(ICHEP2016)434}}
}
\vskip 0.6in

\begin{center}

{\Large {\bf Higgs physics at the Future Circular Collider}}

\vskip 0.2in

{\bf David d'Enterria}$^{1}$

\vskip 0.1in

{\small {\it $^{1}$CERN, EP Department, 1211 Geneva, Switzerland}}

\vskip 0.2in

{\bf Abstract}

\end{center}

\baselineskip=18pt \noindent

%%%%%%%%%%%%%%%%%%%%%%%%%%%%%%%%%%%%%%%%%%%%%%%%%

\noindent The unique Higgs physics opportunities accessible at the CERN Future Circular Collider (FCC) 
in electron-positron ($\sqrts$~=~125,~240, 350~GeV) and proton-proton ($\sqrts$~=~100~TeV) collisions,
are succinctly summarized. Thanks to the large \cm\ energies and enormous luminosities (plus 
clean experimental conditions in the $\epem$ case), many open fundamental aspects of the Higgs 
sector of the Standard Model (SM) can be experimentally studied:
\begin{description}
\item (i) Measurement of the Higgs Yukawa couplings to the lightest fermions: u,d,s quarks
(via rare exclusive $\rm H\to(\rho,\omega,\phi)+\gamma$ decays); and e$^\pm$ (via resonant s-channel $\epem\!\to\,$H production); 
as well as neutrinos (within low-scale seesaw mass generation scenarios).
\item (ii) Measurement of the Higgs potential (triple $\lambda_3$, and quartic $\lambda_4$ self-couplings), via 
double and triple Higgs boson production in pp collisions at 100 TeV.
\item (iii) Searches for new physics coupled to the scalar SM sector at scales $\Lambda_{_{\rm NP}}\gtrsim$~6~TeV, 
%as a means to solve the fine tuning problem, 
thanks to measurements of the Higgs boson couplings with subpercent uncertainties in $\epem\!\to H\,Z$.
\item (iv) Searches for dark matter in Higgs-portal interactions, via high-precision measurements of 
on-shell and off-shell Higgs boson invisible decays.
\end{description}
All these measurements are beyond the reach of pp collisions at the Large Hadron Collider. New higher-energy $\epem$ 
and pp colliders such as FCC are thus required to complete our understanding of the full set of SM Higgs parameters,
%of the SM Higgs sector, as well as to search for new physics coupled to it in the multi-TeV regime.
as well as to search for new scalar-coupled physics in the multi-TeV regime.

%%%%%%%%%%%%%%%%%%%%%%%%%%%%%%%%%%%%%%%%%%%%%%%%

%\vskip 0.8in

%\leftline{January 2017}

\end{titlepage}

\newpage

%\scrollmode

%%%%%%%%%%%%%%%%%%%%%%%%%%%%%%%%%%%%%%%%%%%%%%%%%%%%%%%%%%%%%%%%%%%%%%%%%%%%%%%%%%%%%%%%%%%%%%%%%%%%%%%%%%%%%%%%%%%%
\section{Introduction}

Despite its tremendous success describing many phenomena with high accuracy ---crowned with the discovery of 
its last missing piece, the Higgs boson, in 2012~\cite{Aad:2012tfa,Chatrchyan:2012xdj}--- many fundamental 
questions of the Standard Model (SM) of particle physics still remain open today. Our lack of understanding 
of the nature of dark matter, the origin of matter-antimatter asymmetry, the generation of neutrino masses, 
or how to tame the quadratically-divergent virtual SM corrections affecting the running of the Higgs boson 
mass between the widely separated electroweak and Planck scales (``fine tuning'' problem), 
among others, are questions which likely will {\it not} be fully answered through the study of 
proton-proton (pp) collisions at the Large Hadron Collider (LHC). Searching for solutions to such
fundamental problems, together with a {\it complete} experimental confirmation of the SM Higgs sector 
---including the unknown Yukawa couplings of the lightest fermions, as well as the triple $\lambda_3$ 
and quartic $\lambda_4$ Higgs self-couplings--- requires both a new pp collider at higher center-of-mass (\cm)
energies, as well as a new high-precision $\epem$ machine with unprecedented luminosities to very accurately
study the H boson properties. The Future Circular Collider (FCC) is a post-LHC project in a new 100-km 
tunnel under consideration at CERN~\cite{Benedikt:2015poa}, designed to deliver pp at 
$\sqrts$~=~100~TeV with $\Lint$~=~0.2--2~ab$^{-1}$/yr integrated luminosities (FCC-hh)~\cite{Zimmermann:2016puu}, 
as well as $\epem$ over $\sqrts$~=~90--350~GeV with up to 80~ab$^{-1}$/yr (FCC-ee)~\cite{TLEP}.
Both machines are truly competitive ``Higgs factories''.
\begin{figure}[htbp!]
\centering
\includegraphics[width=0.56\columnwidth,height=5.75cm]{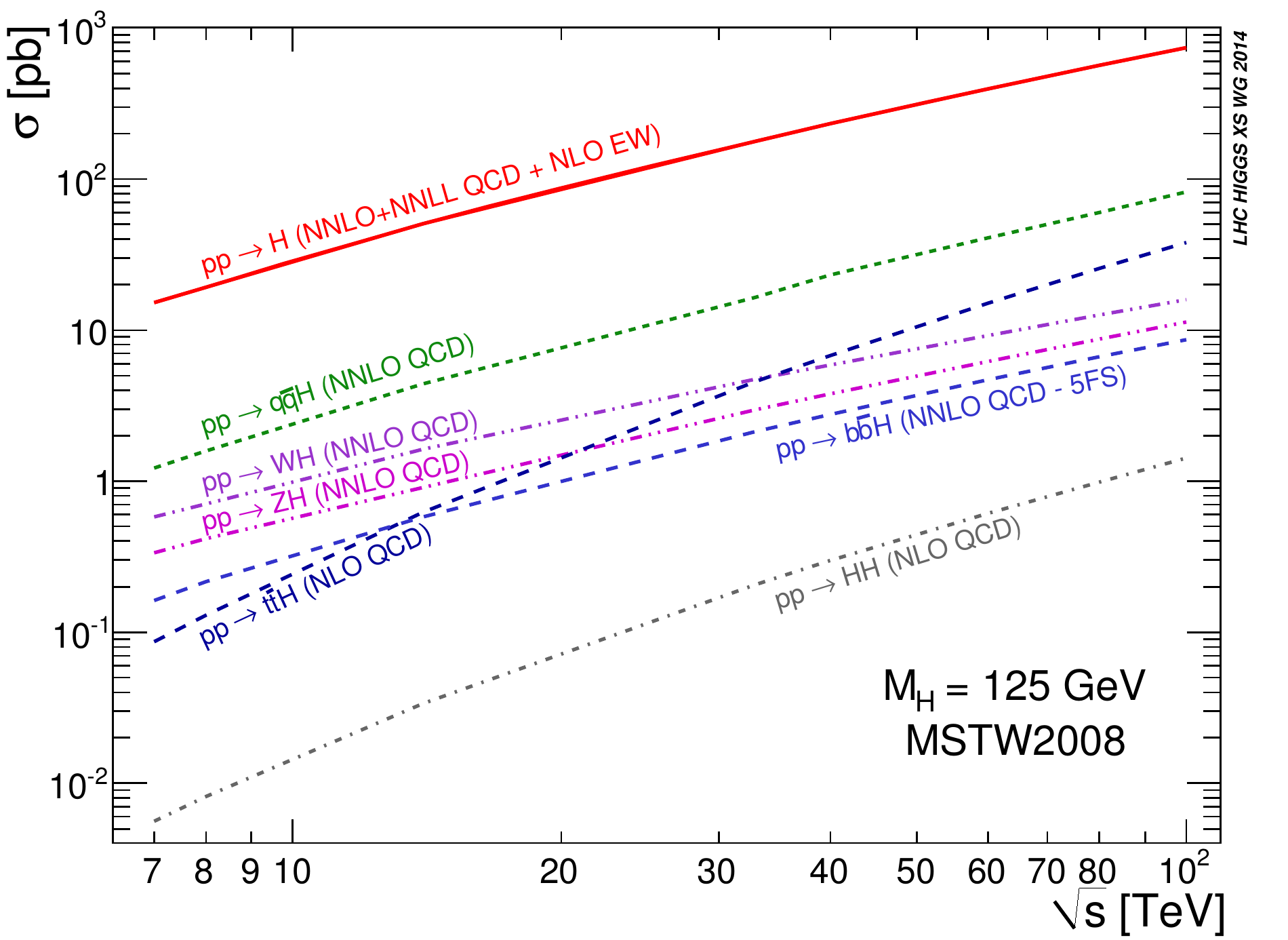}%\hspace{0.2cm}
\includegraphics[width=0.44\columnwidth,height=5.5cm]{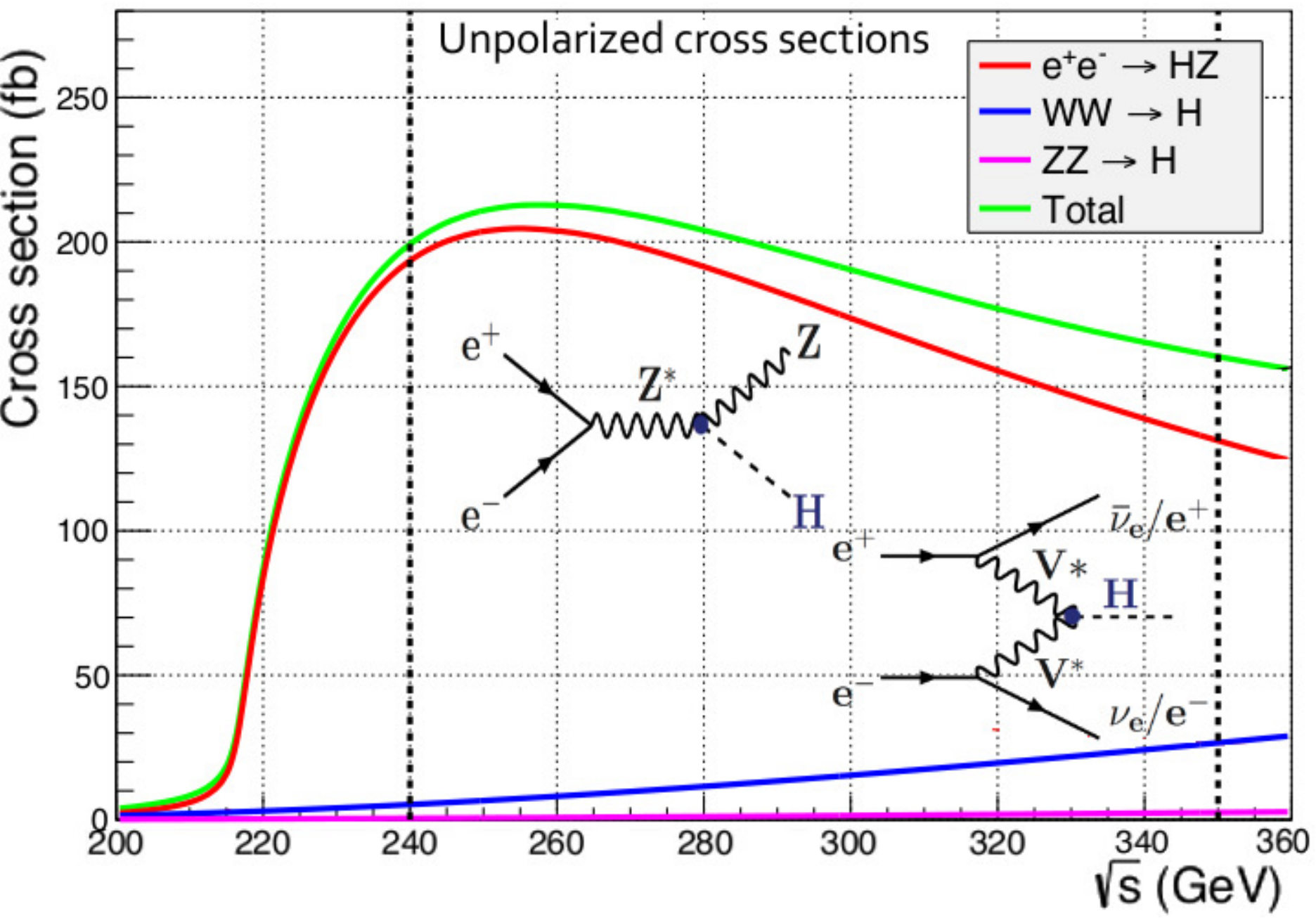}
\caption{Higgs boson cross sections as a function of \cm\ energy (total, and separated for different subprocesses)
in pp  (left)~\cite{Contino:2016spe} and $\epem$ (right)~\cite{TLEP} collisions.}
\label{fig:sigmaH}
\end{figure}
Figure~\ref{fig:sigmaH} shows the H boson production cross sections as a function of \cm\ energy at FCC-hh 
(left) and FCC-ee (right). Higgs production in DIS %deep-inelastic scattering 
at FCC-eh, not discussed here, 
is also possible~\cite{Klein:2016uwv}. At the FCC-hh, gluon-gluon fusion (ggF) dominates the cross section, followed by 
vector-boson-fusion (VBF), and associated $\ttbar$ production. 
The sum of all contributions amounts to $\rm \sigma(pp\to H+X)\approx$~0.9~nb at $\sqrts$~=~100~TeV~\cite{Contino:2016spe}. 
At the FCC-ee, the cross section (rates) peaks at $\rm \sigma(\epem\!\to H+X)\approx$~200~fb at $\sqrts\approx$~240~(250)~GeV~\cite{TLEP}, 
dominated by Higgsstrahlung ($\epem\!\to{\rm HZ}$) with small VBF contributions ($\rm VV\to {\rm H}\;\epem,\nu\nu$). 
Both machines provide unparalleled opportunities to study the Higgs sector of the SM thanks to the 
enormous number of scalar bosons produced over $\sim$15 and $\sim$20 years of operation:
up to 2$\cdot 10^{6}$ at FCC-ee with very low backgrounds and no pileup, and 2$\cdot 10^{10}$ at FCC-hh. 
Measurements of very precise Higgs couplings (with subpercent uncertainties), and of very rare and beyond 
the SM decays are thereby possible.

%%%%%%%%%%%%%%%%%%%%%%%%%%%%%%%%%%%%%%%%%%%%%%%%%%%%%%%%%%%%%%%%%%%%%%%%%%%%%%%%%%%%%%%%%%%%%%%%%%%%%%%%%%%%%%%%%%%%
\section{Generation of the lightest fermion (u, d, s; e; and $\nu$'s) masses}

The SM Higgs boson couples to the fundamental fermions proportionally to their masses $\rm m_f$, and thus
%(or to the square boson masses $\rm m_V^2$). the Higgs boson 
its decays into the actual constituents of the stable visible matter in the Universe
---formed by first  %and second ($\ssbar$) 
generation fermions ($\uubar$, $\ddbar$, $\rm e^\pm$) with light masses 
$\cO{\rm 0.5-10~MeV}$-- %$\rm 0.5-100~MeV$, 
have extremely reduced branching ratios and cannot be directly measured 
at the LHC. The large and clean Higgs boson samples at FCC-ee will allow the measurements of very 
rare exclusive decays into light vector-mesons (VM) plus a photon ($\rm H \to \rho, \omega, \phi \,+\gamma$, 
with $\rho = (\uubar-\ddbar)/\sqrt{2}, \omega = (\uubar+\ddbar)/\sqrt{2}, \phi = \ssbar$)
that are sensitive to the lightest quarks' Yukawas. 
The branching ratios for such processes are $\cO{10^{-5}-10^{-6}}$~\cite{Contino:2016spe,Perez:2015lra}. The most promising one is 
$\rm H\to\rho(\pi\pi)\gamma$, with about 40 counts expected with low backgrounds. Determining the 
corresponding sensitivity to the u/d quark Yukawa couplings requires dedicated studies given that the indirect
$\rm H\to \gamma\,\gamma^*\to VM+\gamma$ decays interfere with the direct $\rm H\to VM+\gamma$ ones, 
and dilute the sensitivity to the latter. Of course, all these channels will be produced much more abundantly 
at FCC-hh, but the huge QCD (and pileup) backgrounds jeopardize a possible extraction of the corresponding u,d,s Yukawa
couplings.

\begin{wrapfigure}{r}{0.60\columnwidth}
\centering
%\hspace*{-0.4cm}
\includegraphics[width=1.0\columnwidth]{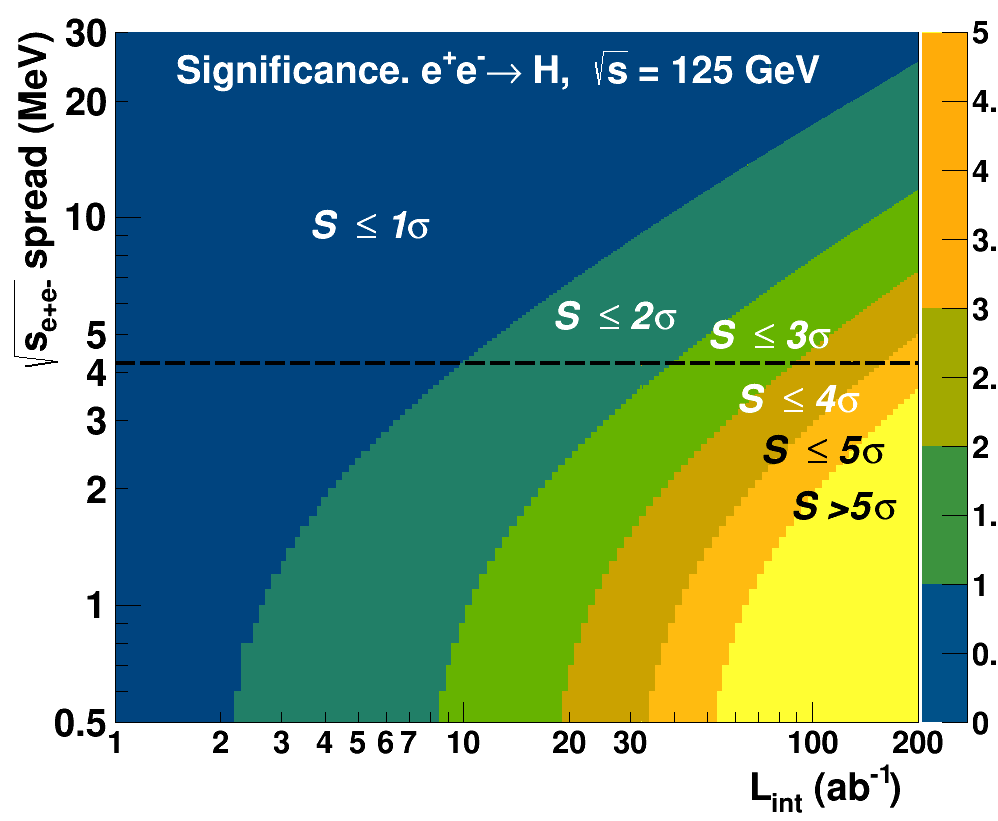}
%\vspace*{-0.5cm}
\caption{Significance contours for the $\epem\!\to$H observation at $\sqrts=125$~GeV 
(combining 10 Higgs boson decays) in the $\sqrts$-spread \vs\ $\Lint$ plane 
at FCC-ee~\cite{DdE_eeH}. The dashed line shows the natural H boson width.}
\label{fig:1stgen}
\end{wrapfigure}

\vspace{0.2cm}
Measuring the electron Yukawa is even harder given the $\rm e^\pm$ lightness, and the only direct method to extract it is 
through resonant s-channel $\epem$ production running at the Higgs pole mass~\cite{DdE_eeH}. The resonant cross section for 
a 125-GeV scalar of natural width $\rm \Gamma_H$~=~4.1~MeV is tiny, $\sigma(\rm \epem\!\to H)$~=~1.64~fb.
The actual cross section is further reduced accounting for the finite energy spread and
initial state photon radiation (ISR) of the $\rm e^\pm$ beams. For a \cm\ energy spread 
commensurate with the $\rm \Gamma_H$ natural width (dashed line in Fig.~\ref{fig:1stgen}), reachable using 
monochromatization~\cite{monochrom}, the cross section becomes $\sigma(\rm \epem\!\to H)$~=~290~ab~\cite{Jadach:2015cwa}. 
Under these conditions, a preliminary study based on counting the number of events for signal and backgrounds 
in 10 different decay final-states %in ten Higgs decay channels 
in $\epem$ at $\sqrts$~=~125.000~$\pm$~0.004~GeV, indicates that a
$3\sigma$ observation requires $\Lint\approx$~90~ab$^{-1}$ (Fig.~\ref{fig:1stgen})~\cite{DdE_eeH}. 
%$3\sigma$ observation requires $\Lint$~=~$\cO{\rm 90~ab^{-1}}$ (Fig.~\ref{fig:1stgen})~\cite{X}. 
For the target $\Lint$~=~40~ab$^{-1}$/yr at 125~GeV, the significance of the signal is 2.1$\sigma$ which 
translates into limits on the $\rm H\to\epem$ branching ratio at $\times$1.2 the SM expectation or, 
equivalently, a 95\% CL upper bound on $\times$1.1 the SM prediction for the $\rm e^\pm$ Yukawa~\cite{DdE_eeH}.\\
%For a more modest $\Lint$~=~10~ab$^{-1}$, the significance of the signal is 0.7$\sigma$ which 
%translates into limits on the $\rm H\to\epem$ branching ratio at $\times$2.8 the SM expectation or, 
%equivalently, a 95\% CL upper bound on $\times$1.7 the SM prediction for the $\rm e^\pm$ Yukawa~\cite{DdE_eeH}.

The generation of non-zero neutrino masses, called for by the observation of their flavor
oscillations, is beyond the SM and requires new particles such as right-handed ``sterile'' $\nu$'s.
Phenomenologically-attractive scenarios have been considered~\cite{Antusch:2016ejd} where sterile 
neutrinos $\rm N_i$ have masses around the electroweak scale, and thereby can be produced at FCC-ee
%in $\epem$ collisions, subsequently 
and observed via $\rm N_i \to H+\nu$. Through the experimental study of mono-Higgs 
final states, FCC-ee has competitive sensitivities for $\rm m_{N_i}\approx$~100--350~GeV and 
values of the active-sterile mixing parameter down to $|\theta_e|^2\approx$~10$^{-5}$.

%%%%%%%%%%%%%%%%%%%%%%%%%%%%%%%%%%%%%%%%%%%%%%%%%%%%%%%%%%%%%%%%%%%%%%%%%%%%%%%%%%%%%%%%%%%%%%%%%%%%%%%%%%%%%%%%%%%%
\section{Determination of the Higgs potential (triple and quartic self-couplings)}

The Higgs sector of the SM cannot be considered to be fully confirmed experimentally until the strength of the 
Higgs boson to itself is measured. The SM Lagrangian parametrizes the Higgs self-interaction
through its triple ($\lambda_3$) and quartic ($\lambda_4$) self-couplings, and their determination is crucial 
to confirm the shape of the Higgs potential and the mechanism of electroweak symmetry breaking~\cite{Djouadi:2005gi}. 
Their direct determination is only possible through the production cross sections of two and three Higgs bosons. 
At the LHC(14 TeV) and FCC(100 TeV), the cross sections amount to $\rm\sigma(HH)\approx$~0.05,\,1.9~pb 
and $\rm\sigma(HHH)\approx$~0.1,\,5~fb~\cite{Contino:2016spe}. However, different production subchannels 
contribute to the HH and HHH cross sections that do not directly involve H self-couplings, thereby 
diluting the final sensitivity on $\lambda_{3,4}$. %Such processes are very difficult or innaccessible at the LHC: 
At the end of the high-luminosity LHC running (HL-LHC, 14~TeV, $\Lint$~=~3~ab$^{-1}$), the uncertainties on $\lambda_3$ 
will be of the order of 50\%~\cite{Dawson:2013bba}, whereas the measurement of $\lambda_4$ is %completely 
out of reach.

\begin{wraptable}{r}{0.55\columnwidth}
\begin{tabular}{l}
\begin{minipage}{\textwidth}
\includegraphics[width=\columnwidth]{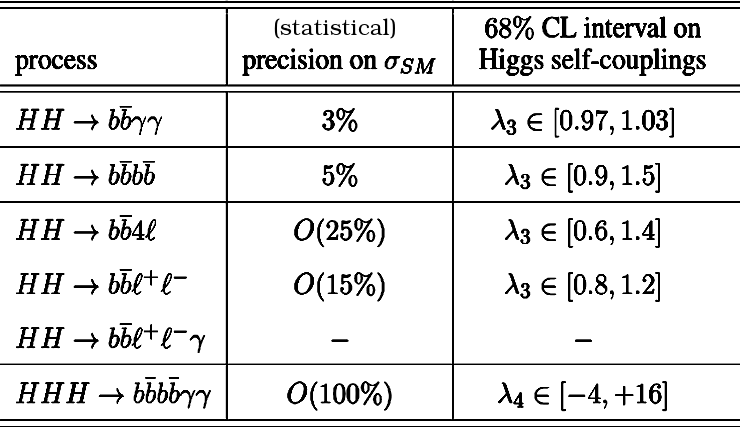}%\hspace{0.2cm}
\end{minipage}
%\vspace*{-0.6cm}
\caption{\label{tab:HH_HHH}Expected precision on SM cross sections for double and triple Higgs final-states 
reachable at FCC-hh (pp at 100~TeV, 30~ab$^{-1}$), and associated 68\% CL ranges on $\lambda_3$ 
and $\lambda_4$ Higgs self-couplings. Details are provided in~\cite{Contino:2016spe}.}%
\end{tabular}
\end{wraptable}

The FCC-ee has a sensitivity to the $\lambda_3$ parameter comparable to that of the HL-LHC, through 
the high-precision study of the dominant H+Z cross section which contains a small (energy-dependent) loop contribution involving the Higgs 
self-coupling~\cite{McCullough:2013rea}. However, a definite $\lambda_3$ measurement and constraints on $\lambda_4$, 
require a 100-TeV pp collider such as FCC-hh. The precision achievable in the measurements of double and 
triple Higgs cross sections at FCC-hh, and associated 68\% CL intervals on the $\lambda_3$ and $\lambda_4$ 
self-couplings are listed in Table~\ref{tab:HH_HHH}~\cite{Contino:2016spe}. The trilinear self-coupling 
can be measured with 3\% uncertainties, whereas the quartic will be mildly constrained.

%%%%%%%%%%%%%%%%%%%%%%%%%%%%%%%%%%%%%%%%%%%%%%%%%%%%%%%%%%%%%%%%%%%%%%%%%%%%%%%%%%%%%%%%%%%%%%%%%%%%%%%%%%%%%%%%%%%%
\section{Searches for new scalar-coupled physics} %Solutions to the hierarchy/naturalness problem}

With the Higgs boson discovered, the SM is now theoretically confronted to the hierarchy (aka.
fine tuning or naturalness) problem, whereby quadratically-divergent SM virtual corrections affect 
the running of the Higgs boson mass between the widely separated electroweak and Planck scales.
New particles are required to stabilize such untamed quantum corrections. Since the Higgs boson 
%is (together with the neutrinos) the least known elementary particle, and since it 
couples directly to any massive particle, the presence of any new physics has large chances 
to affect its couplings to the rest of SM particles.
A powerful model-independent method to encode the effect of new physics from higher energies on experimental 
observables, is provided by the SM Effective Field Theory (EFT), which parametrizes possible new physics
via a systematic expansion in a series of higher-dimensional operators composed of SM fields:
${\cal L}_{\rm eff} = {\cal L}_{\rm SM} + \sum^\infty_{d=5}\frac{1}{\Lambda^{d-4}} {\cal L}_{d}$ with 
${\cal L}_{d} = \sum_i c^d_i {\cal O}_i$, and unknown Wilson coefficients $c_i$ generated by decoupled 
new physics beyond the SM. Often, dim-6 operators ${\cal O}_i$ are the only ones considered (the Weinberg 
neutrino-mass is the unique dim-5 operator, and effects of $d>6$ operators are subleading in the decoupling assumption), 
\ie\ ${\cal L}_{\rm eff} = {\cal L}_{\rm SM} + \sum_{i}\frac{c_i}{\Lambda^2}{\cal O}_i$, where $\Lambda$ 
represents the scale of new interactions, and the coefficients $c_i$ depend on the details of its 
structure~\cite{Pomarol:2013zra,Ellis:2015sca}.
%By extending the SM Lagrangian with new dimension-5 and/or 6 operators:
%\begin{equation}
%\rm {\cal L}_{eff} = \sum^\infty_{d=4}\frac{1}{\Lambda^{d-4}} = {\cal L}_{SM} + 
%                    \frac{1}{\Lambda}{\cal L}_{5} + \frac{1}{\Lambda^2}{\cal L}_{6}, 
%\rm{\;\; with \;\;} {\cal L}_{d} = \sum_i c^d_i {\cal O}_i
%\end{equation}
In the case of indirect (loop) constraints on new physics coupled to the Higgs boson, a useful 
back-of-the-envelope formula can be derived which relates $\Lambda$ %the scale of new physics 
to deviations of its couplings ($\delta g_{_{\rm HXX}}$) with respect to the expected SM values:\\
\vspace{-0.2cm}
\begin{equation}
%\Lambda \gtrsim (\rm{1 TeV})/\sqrt{(\delta g_{\mathsc{h_{xx}}}/g_{\mathsc{h_{xx}}})/5\%};
%\[
%\begin{center}
\Lambda \gtrsim (\rm{1~TeV})/\sqrt{(\delta g_{_{HXX}}/g_{_{HXX}})/5\%};
%\end{center}
%\vspace{-0.5cm}
\label{eq:Lambda_H}
\end{equation}
\ie\ H couplings measurements of 5\% precision are sensitive to new physics at $\Lambda\gtrsim$~1~TeV.\\
%\vspace{-0.3cm}

\begin{table}[htbp!]
%\begin{wraptable}{r}{0.6\columnwidth}
%\scriptsize
\renewcommand{\arraystretch}{1.15}
\begin{tabular}{l|cc|cccc}\hline\hline
Parameter                   &Current$^*$&HL-LHC$^*$& FCC-ee & ILC    & CEPC & CLIC \\
                            &7+8+13 TeV & 14 TeV & Baseline & Lumi upgrade & Baseline & Baseline \\ %0.35+1.4+3 TeV \\
                            & $\cO{\rm 70~fb^{-1}}$ & (3 ab$^{-1}$) & (10 yrs) & (20 yrs) & (10 yrs) & (15 yrs) \\\hline  
$\rm \sigma(HZ)$            & --    &  --          & 0.4\%  & 0.7\%  & 0.5\% & 1.6\% \\
$\rm g_{_{ZZ}}$             & 10\%  &  2--4\%      & 0.15\%  & 0.3\% & 0.25\% & 0.8\% \\
$\rm g_{_{WW}}$             & 11\%  &  2--5\%      & 0.2\%  & 0.4\%  & 1.6\% & 0.9\% \\\hline
$\rm g_{_{bb}}$             & 24\%  &  5--7\%      & 0.4\%  & 0.7\%  & 0.6\% & 0.9\% \\
$\rm g_{_{cc}}$             & --    &  --          & 0.7\%  & 1.2\%  & 2.3\% & 1.9\% \\
$\rm g_{_{\tau\tau}}$       & 15\%  &  5--8\%      & 0.5\%  & 0.9\%  & 1.4\% & 1.4\% \\
$\rm g_{_{\ttbar}}$         & 16\%  &  6--9\%      & 13\%   & 6.3\%  & -- & 4.4\% \\
$\rm g_{_{\mu\mu}}$         & --    &  8\%         & 6.2\%  & 9.2\%  & 17\%  & 7.8\% \\
%$\rm g_{_{\epem}}$          & --    &  --          & $<$100\%~\cite{DdE_eeH} & --     & --    & -- \\\hline
$\rm g_{_{\epem}}$          & --    &  --          & $<$100\% & --     & --    & -- \\\hline
$\rm g_{_{gg}}$             & --    &  3--5\%      & 0.8\%  & 1.0\%  & 1.7\% & 1.4\% \\
$\rm g_{_{\gamma\gamma}}$   & 10\%  &  2--5\%      & 1.5\%   & 3.4\%  & 4.7\% & 3.2\% \\
$\rm g_{_{Z\gamma}}$        & --    &  10--12\%    & \multicolumn{3}{c}{(to be determined)} & 9.1\% \\\hline
$\rm \Delta m_{_{H}}$  & 200 MeV & 50 MeV             & 11 MeV     & 15 MeV     & 5.9 MeV & 32 MeV \\
$\rm \Gamma_{_{H}}$   & $<$26 MeV& 5--8\%          & 1.0\%  & 1.8\% & 2.8\% & 3.6\% \\
$\rm \Gamma_{_{inv}}$            & $<$24\%&  $<$6--8\%& $<$0.45\%& $<$0.29\%& $<$0.28\% & $<$0.97\% \\
\hline\hline
\end{tabular}
\caption{\label{tab:H}Summary of the best statistical precision attainable for Higgs observables at future $\epem$ colliders
(FCC-ee~\cite{TLEP}, ILC~\cite{ILC}, CEPC~\cite{CEPC}, CLIC~\cite{CLIC}) compared to (model-dependent$^*$)
current LHC~\cite{Khachatryan:2016vau} and expected HL-LHC~\cite{Dawson:2013bba} 
pp results.}%
%\end{wraptable}
\end{table}
%\noindent 

At lepton colliders, precise and model-independent Higgs measurements can be carried out 
using the recoil mass method in $\epem\!\to\rm HZ$, which allows an accurate determination of 
the H boson 4-momentum irrespective of its decay mode, from the Z$\to\ell^+\ell^-$ ($\rm \ell=e,\mu$) 
decay reconstruction. %(Fig.~\ref{fig:Higgs}, left). 
At the FCC-ee, the high-precision ($\pm$0.4\%) measurement of $\sigma_{\rm \epem\!\to HZ}\propto g_{_{\rm HZ}}^2$ 
provides a {\it model-independent} value of $g_{_{\rm HZ}}$ to within $\pm$0.2\%. 
%and, according to Eq.~(\ref{eq:Lambda_H}), $\Lambda\gtrsim$~5~TeV. 
The total Higgs boson width $\Gamma_{\rm H}$ can also be obtained with 1\% uncertainty combining the 
measured value of $\sigma_{\rm \epem\!\to H(XX)Z}\propto\Gamma_{\rm H\to XX}$ with the known branching 
fractions, BR$_{\rm X}=\Gamma_{\rm H \to XX}/\Gamma_{\rm H}$, for different decays.
The Higgs mass can be determined to within $\pm$11~MeV from the measured recoil mass. 
%in the Z$\to\ell^+\ell^-$ ($\rm \ell=e,\mu$) channel. 
Table~\ref{tab:H} provides a summary of the {\it best} precision attainable for most Higgs boson 
properties at future $\epem$ machines (FCC-ee~\cite{TLEP}, ILC~\cite{ILC}, CEPC~\cite{CEPC}, and CLIC~\cite{CLIC}) 
compared to those today~\cite{Khachatryan:2016vau} and reachable at HL-LHC~\cite{Dawson:2013bba}. 
Lepton colliders provide a factor of at least 50 (10) improvement with respect to the present (HL-LHC) 
results that, in addition and at variance with the latter, do not depend on any SM fit. Among future 
$\epem$ colliders, FCC-ee typically features the highest precision thanks to its expected higher luminosities.
Farther, the 2$\cdot 10^{10}$ scalars bosons produced at FCC-hh will also systematically improve the precision 
of all H couplings, preliminary studies~\cite{Plehn:2015cta,Contino:2016spe} indicate a potential precision of 
1\% for those with lower rates at $\epem$ machines: $g_{\ttbar}$, $g_{_{\mu\mu}}$, and $g_{_{\rm Z\gamma}}$.

\begin{figure}[htbp!]
\centering
\includegraphics[width=0.5\columnwidth]{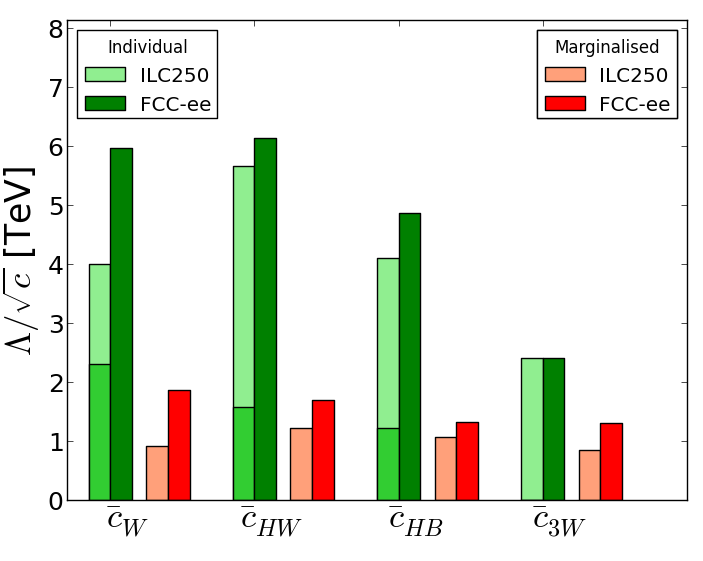}%\hspace{0.2cm}
\includegraphics[width=0.5\columnwidth,height=6.1cm]{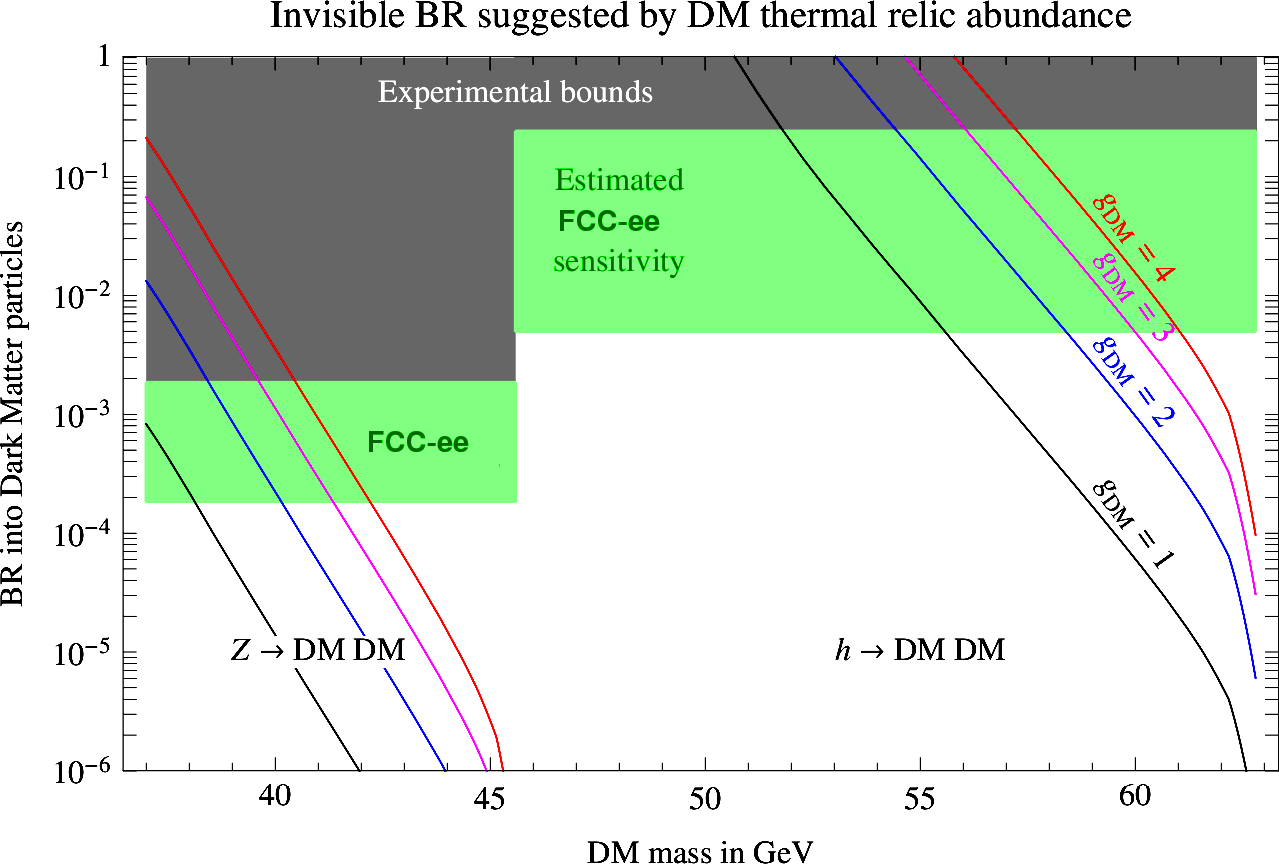}%\hspace{0.1cm}
\caption{Left: Sensitivity reach to new physics scales ($\Lambda/\sqrt{c_i}$), encoded in four dim-6 operator $c_i$ coefficients,
of precision Higgs (and triple gauge boson couplings) measurements at FCC-ee and ILC~\cite{Ellis:2015sca}.
Right: FCC-ee sensitivity for rare H (and Z) decays into DM pairs in the $\rm BR_{_{H,Z\to\phi\phi}}$ \vs\ 
$\rm m_{\phi}$ plane~\cite{deSimone:2014pda}.}
\label{fig:FCCee_H}
\end{figure}

The most precise coupling at FCC-ee ($\delta \rm g_{_{ZZ}}/g_{_{ZZ}}\approx$~0.15\%) will allow setting 
limits on new scalar-coupled physics at $\Lambda\gtrsim$~5.8~TeV as per the simple estimate~(\ref{eq:Lambda_H}). 
Accurate theoretical analyses based on dim-6 EFT~\cite{Ellis:2015sca,deBlas:2016ojx} yield indeed $\Lambda\gtrsim$~6~TeV 
(Fig.~\ref{fig:FCCee_H}, left).
%The precision Higgs studies at the FCC-ee not only impose generic constraints on new physics scales at multi-TeV energies, 
The same measurements can also be interpreted in terms of sensitivity to broad classes of SUSY models (such as the Constrained MSSM, 
%Non-Universal Higgs Masses
or natural SUSY) effectively covering phase space corners beyond the LHC 
reach~\cite{Buchmueller:2015uqa,Fan:2014axa}. %Figure~\ref{fig:susy} (left) 
%compares the precision Higgs observables at the LHC, ILC and FCC-ee with the deviations from their SM values
%expected for the low- and high-mass CMSSM and NUHM1 best-fit points~\cite{TLEP,Buchmueller:2015uqa}, emphasizing
%the strong FCC-ee ability to distinguish these models from the SM. The high-mass CMSSM points in Fig.~\ref{fig:susy},
%provide, in particular, typical SUSY scenarios which likely lie beyond the LHC reach featuring narrow strips 
%where stop-neutralino coannihilation is important, or focus-point strips at higher values of the ratio $m_0/m_{1/2}$. 
%The Higgs measurements at the FCC-ee would be able to probe both types of narrow parameter-space strips that extend 
%to large sparticle masses, and indirectly determine CMSSM parameters also in such a pessimistic scenario~\cite{Buchmueller:2015uqa}.

%%%%%%%%%%%%%%%%%%%%%%%%%%%%%%%%%%%%%%%%%%%%%%%%%%%%%%%%%%%%%%%%%%%%%%%%%%%%%%%%%%%%%%%%%%%%%%%%%%%%%%%%%%%%%%%%%%%%
\section{Searches for Higgs-portal dark matter (DM)}

The SM describes only 4\% of the universe energy budget, the rest being in the form of unknown 
DM (and dark energy) contributions, pointing to the existence of  new massive particles (such as \eg\ SUSY partners, 
heavy $\nu$'s, axions,...). In Higgs-portal models~\cite{Djouadi:2012zc}, the H boson acts as a mediator between 
the SM and DM particles, playing a central role in the evolution of the early universe. Attractive scenarios exist for 
DM candidates ($\phi$) lighter than $m_{_{\rm H,Z}}/2$, consistent with the measured DM thermal relic abundance in the universe, 
with DM freezing out through resonant H (or Z) exchanges. In such cases, the measurements of the invisible H and 
Z widths %at the FCC-ee 
provide the best collider options to test such scenarios~\cite{deSimone:2014pda}. 
Current invisible H decays limits are BR(H$\to$inv.)$\,<$~0.24 (95\% CL) at the LHC~\cite{Khachatryan:2016whc}, and are 
expected to reach BR(H$\to$inv.)~$\!<$~0.06 at HL-LHC (Table~\ref{tab:H}). At the FCC-ee, the $\rm H\,Z(\ell^+\ell^-)$ 
final state can be used to directly measure $\Gamma_{\rm inv}$ %the Higgs invisible width 
(a 5$\sigma$ observation is possible down to BR~=~1.7$\pm$0.1\%)~\cite{Cerri:2016bew}, %with an absolute precision below 0.2\%, 
in events where its decay products escape undetected. If unobserved, a 0.5\% upper limit (95\% CL)~\cite{TLEP} can be 
set on this branching ratio (Fig.~\ref{fig:FCCee_H}, right), placing DM bounds a factor of 
50 (10) better than those at LHC (HL-LHC), and being also competitive with the reach of planned direct 
detection experiments for $\rm m_{\phi}<$~10~GeV~\cite{Cerri:2016bew}.\\
\vspace{-0.3cm}

For DM particles heavier than the Higgs boson, off-shell H decays into DM can be searched in pp %at FCC-hh 
events characterized by missing energy (from the H$^\star\to\phi\phi$ decay) accompanied by extra particle 
production (gluon ISR in ggF, forward-backward jets in VBF, associated $\ttbar$,...) as done at the LHC~\cite{Khachatryan:2016vau}. 
Theoretical studies indicate that the FCC-hh can place strong constraints on Higgs-portal couplings 
$|c_\phi|\approx$~0--3.5 for scalar DM masses $m_\phi$~=~150--500~GeV (see details in~\cite{Contino:2016spe,Craig:2014lda}).

%%%%%%%%%%%%%%%%%%%%%%%%%%%%%%%%%%%%%%%%%%%%%%%%%%%%%%%%%%%%%%%%%%%%%%%%%%%%%%%%%%%%%%%%%%%%%%%%%%%%%%%%%%%%%%%%%%%%
\section{Summary}

\noindent
The Future Circular Collider (FCC) will provide unparalleled luminosities $\cO{\rm 1-20~ab^{-1}}$/yr in
pp ($\sqrts$~=~100~TeV) and $\epem$  ($\sqrts$~=~125--350~GeV) collisions, totalling 2$\cdot 10^{10}$ and 
2$\cdot 10^{6}$ Higgs bosons produced over their respective expected operation times, and opening up %the possibility to carry out 
measurements with $\cO{50}$ ($\cO{10}$) times better precision than those reachable at LHC (HL-LHC).
%beyond the reach of any other current or future experimental facility. 
The unique FCC Higgs physics opportunities include fully closing the SM scalar sector (measuring the unknown 
Yukawas of the first-generation fermions, as well as the triple and quartic Higgs self-couplings), 
and discovering (or placing bounds on) scalar-coupled new physics well into the multi-TeV regime.\\
%and improving (by factors $\cO{50}$ and $\cO{10}$ with respect to LHC and HL-LHC) 
%our sensitivity to scalar-coupled new physics well into the multi-TeV regime.

%%%%%%%%%%%%%%%%%%%%%%%%%%%%%%%%%%%%%%%%%%%%%%%%%%%%%%%%%%%%%%%%%%%%%%%%%%%%%%%%%%%%%%%%%%%%%%%%%%%%%%%%%%%%%%%%%%%%

\noindent{\bf Acknowledgments--\;} Discussions (and/or feedback to a previous version of this document) 
with A.~David, C.~Grojean, P.~Janot, M.~Mangano, and J.~Tian are gratefully acknowledged. %

\end{document}